\documentclass[12p]{article}
\usepackage{latexsym}
\usepackage{color}
\usepackage{amsmath}
\usepackage{epsfig}
\usepackage{hyperref}
 \usepackage{subfigure}
 \usepackage{dcolumn}
   \usepackage{threeparttable}

\newcommand{\ortala}[1]{\begin{center}#1\end{center}}

\newcommand{\ket}[1]{\left|#1\right\rangle}

\newcommand{\sand}[3]{\left\langle #1\left|#2\right|#3\right\rangle}
\newcommand{\sandd}[1]{\left\langle #1\right\rangle}

\newcommand{\summ}[3]{{{\underset{#1 }{\overset{#2}{\displaystyle\sum}}}#3}}
\newcommand{\prodd}[3]{{{\underset{#1
}{\overset{#2}{\displaystyle\prod}}}#3}}

\newcommand{\re}[1]{(\ref{#1})}

\newcommand{\eq}[2]{\begin{equation}\label{#1}  #2\end{equation}}

\newcommand{\paran}[1]{\left(#1\right)}

\newcommand{\sch}[1]{Schrodinger}

\newcommand{\komb}[2]{\paran{\begin{array}{c} #1 \\ #2 \end{array}}}

\newenvironment{changemargin}[2]{%
\begin{list}{}{%
\setlength{\topsep}{0pt}%
\setlength{\leftmargin}{#1}%
\setlength{\rightmargin}{#2}%
\setlength{\listparindent}{\parindent}%
\setlength{\itemindent}{\parindent}%
\setlength{\parsep}{\parskip}%
}%
\item[]}{\end{list}}

\usepackage{amssymb}
\usepackage{latexsym}
\begin{document}

\ortala{\large\textbf{Anisotropic Heisenberg Model on Core-Shell Structured Nanotube Geometry}}

\ortala{\textbf{\"Umit Ak\i nc\i \footnote{\textbf{umit.akinci@deu.edu.tr}}}}

\ortala{\textit{Department of Physics, Dokuz Eyl\"ul University,
TR-35160 Izmir, Turkey}}

\section{Abstract}

The effect of the anisotropy in the exchange interaction on the critical temperature and the order parameter
of the anisotropic Heisenberg model on a core-shell structured nanotube geometry
has been investigated.  As a formulation, effective field theory with the differential
operator technique and decoupling approximation within the 4-spin cluster
(EFT-4) has been used. The variation of the critical temperature with the
anisotropy in the exchange interaction has been obtained. Besides, suitability of the effective field theory 
in small clusters as a formulation for the Heisenberg nanotube has been discussed and it is concluded that 
minimum size of the cluster is 4-spin cluster, which can give the correct critical description of the system.

\section{Introduction}\label{introduction}
Magnetic nanomaterials promise a wide variety of technological applications because they have
a great many unusual and interesting thermal and magnetic behaviors. Due to the recent
developments in experimental techniques permit us to produce different types of nanomaterials
with a few atoms such as nanowires, nanotubes, nanorods and nanocubes. As a result of these
outstanding properties, there has been growing interest in the study of magnetic nanomaterials
both theoretically and experimentally. These nanomaterials have been already used with a
great success in many different areas such as sensors \cite{ref1}, permanent
magnets \cite{ref2} and medical applications  \cite{ref3}. Magnetic properties and phase
transition characteristics of nanoparticles strongly depend on the size and the dimensionality of the particle.

It should be mentioned that nanotubes have been attracting
considerable attention, especially after the discovery of carbon nanotubes \cite{ref4}, and
investigation of the mechanical, electrical, optical and magnetic properties of nanotubes
is still an active research area. At the present time, ferromagnetic  nanotubes
have been successfully fabricated   \cite{ref5, ref6}, and they are
promising for various technological applications \cite{ref7, ref8}.

Various types of nano structures, such as $\mathrm{FePt}$ and $\mathrm{Fe_3O_4}$ nanotubes \cite{ref9},
can be modeled by core-shell models and the physical properties of considered systems can be determined
by benefiting from the well defined methods such as  Mean Field Approximation (MFA),
Effective Field Theory (EFT) and Monte Carlo Simulation Techniques (MC). From the theoretical
point of view, phase transition characteristics as well as other magnetic properties of
core-shell nanotube systems have been determined by making use of classic Ising model within the framework
of standard EFT formulation \cite{ref10, ref11, ref11_ek, ref12, ref13, ref14, ref15}.

As far as we know, there exists a limited number of studies focusing on the physical nature of the
core-shell nanotube system using Heisenberg model. For example, classical Heisenberg model on a
single wall ferromagnetic nanotubes has been solved with MC \cite{ref16, ref17} and many-body Green's
function method \cite{ref18}. Furthermore, three-leg quantum spin tube \cite{ref19} has been solved with numerical
exact diagonalization within the finite-cluster \cite{ref20}.

The aim of the present work is to probe the influences of the anisotropy in the exchange spin-spin
interactions on the critical temperature and the order parameter of the  core-shell structured Heisenberg nanotube in detail.
For this aim, we  use the EFT formulation in 4-spin cluster. It is a well known fact that EFT  can provide
better results than MFA, due to the consideration of self spin correlations, which are omitted
in the MFA. EFT-2 formulation \cite{ref21} mostly used for the Heisenberg model. However,
as discussed  in Sec. \ref{formulation}, aforementioned formulation can not give
satisfactory results for this system. Hence, we derive EFT formulation for 4-spin cluster for the
core-shell nanotube system.

The paper is organized as follows: In Sec. \ref{formulation} we
briefly present the model and  formulation, the results and discussions are presented in Sec. \ref{results},
and finally Sec. \ref{conclusion} contains our conclusions. We have also two
appendices for the details of the formulation.

\section{Model and Formulation}\label{formulation}

\begin{figure}[h]\begin{center}
\epsfig{file=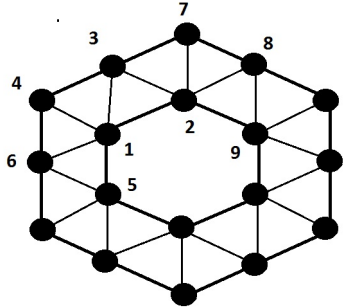, width=10cm}
\end{center}
\caption{Schematic representation of the nanotube (top view).} \label{sek1}\end{figure}

One layer of the nanotube system is given in Fig. \ref{sek1}. The system is infinitely long along
the  axes which is perpendicular to the figure plane. The inner portion
of the system called core (number of 6 spins in each layer), while the
outer one called shell which has number of 12 spins in each layer.

\noindent The Hamiltonian of the anisotropic Heisenberg model can be  simply given by
\eq{denk1}{\mathcal{H}=-\summ{<i,j>}{}{\paran{J_{ij}^x S_i^xS_j^x+J_{ij}^y S_i^yS_j^y+
J_{ij}^z S_i^zS_j^z}},} where $S_i^x,S_i^y$ and  $S_i^z$ denote the Pauli spin operators
at a site $i$. $J_{ij}^x,J_{ij}^y$ and $J_{ij}^z$ stand for the components of the exchange
interactions between the nearest neighbor spins located at sites $i$ and $j$. The sum
is carried over the nearest neighbors of the lattice. The exchange interaction
components $(J_{ij}^x,J_{ij}^y,J_{ij}^z)$ between the spins on the sites $i$ and $j$ takes
the values according to the positions of these nearest neighbor spins. Let the exchange
interaction components between the nearest neighbor spins located at the core
be $(J_{1}^x,J_{1}^y,J_{1}^z)$ and the shell be $(J_{2}^x,J_{2}^y,J_{2}^z)$. The exchange
interaction components between the nearest neighbor spins located at the core and
the shell be $(J_{3}^x,J_{3}^y,J_{3}^z)$.

As discussed above in short, EFT-2 formulation \cite{ref21} mostly used in
the literature to analyze the phase transition properties of
different types of magnetically interacting many-body systems. It is possible to say that
it is generalization of the same formulation within the Ising model \cite{ref22} to
the Heisenberg model. Strictly speaking, in EFT-2 formulation,
two spins located $i.$ and $j.$ sites of system (i.e. $S_i$ and $S_j$) are selected
and then the 2-spin cluster is constructed by benefiting from the selected spins.
Interactions between these two spins treated exactly whereas interactions between a
spin in a chosen cluster and a spin belongs to the out of the
cluster treated approximately. Some mathematical difficulties arise due to
nature of the formulation, and in order to avoid it we replace the perimeter
spins of the 2-spin cluster by Ising spins (axial approximation) \cite{ref23}.
The aforementioned formulation has been successfully applied to a great deal number of
geometries  such as bulk materials \cite{ref21}, and finite sizes
magnetic systems \cite{ref24, ref25}. We should also emphasize that the finite-size systems
have multiple exchange interactions between different spins, so multiple clusters
are required to define the considered system in a good way. In contrary to the finite-size systems,
one cluster is enough for infinite-size materials since they have
translational invariance properties. One of the ways providing mathematical simplicity
is axial approximation but it brings about some deficiencies. For example, with using
axial approximation the Heisenberg character of the exchange interaction is loss and
this corresponds to working spins which are outside of the cluster with only $z$ component.
It is possible say that one of the two nearest neighbor spins having only $z$ component
is in the selected cluster and the other is outside of the cluster.

It is clear that EFT-2 formulation with axial approximation can not give reasonable results for
the nanotube geometry. The considered system has three different types of exchange interactions
and  2-spin cluster formulation can not handle three of them exactly with the Heisenberg character.
It should be underlined that larger cluster are needed in order to get a better
result for this problem. Indeed, the numerical processes begin to complicate and
computational cost rises as size of the cluster increases. One of the suitable
choices is 4-spin cluster called EFT-4 formulation, and it has been already applied to the
investigation of Ising model \cite{ref26}. This kind of cluster can be constructed with four
spins  labeled by $1, 2, 3$ and $4$ in Fig. \ref{sek1}.

The 4-spin Hamiltonian of the selected cluster is as following
\eq{denk2}{\mathcal{H}^{(4)}=-\summ{<i,j>}{}{}\paran{J_{ij}^x S_i^xS_j^x+
J_{ij}^y S_i^yS_j^y+J_{ij}^z S_i^zS_j^z}-\summ{i=1}{4}{}h_iS_i^z.}
The first sum is carried over the nearest neighbor spins inside the chosen cluster,
while the second one is over the spins which are inside  the chosen cluster. The labels of the
spins which are inside the cluster and exchange interactions between them are ($S_1,S_2,J_{1}$),
($S_2,S_3,J_{3}$), ($S_3,S_4,J_{2}$) and ($S_4,S_1,J_{3}$). The terms in the second sum of
Eq. \re{denk2} $h_i (i=1,2,3, 4)$, denotes the all interactions of the spin labeled by $i$
with the nearest neighbor spins, which are outside of the cluster.

Thermal expectation value of any observable quantity which is the function of the spin
components $\Omega$ can be obtained by exact generalized of Callen-Suzuki identity \cite{ref27}
\eq{denk3}{
\sandd{\Omega}=\sandd{\frac{Tr_{4}\Omega \exp{\paran{-\beta \mathcal{H}^{(4)}}}}{Tr_{4}\exp{\paran{-\beta \mathcal{H}^{(4)}}}}}
,} where $Tr_{4}$ means that the partial trace operation over the degrees of the
freedom labeled by $1,2,3,4$, $\beta=1/(k_B T)$, where $k_B$ is Boltzmann constant
and $T$ is the temperature and $\mathcal{H}^{(4)}$ is the Hamiltonian of the 4-spin cluster
within the axial approximation. The matrix representations of the operators located at the numerator
and the denominator of the right hand side of Eq. \re{denk3} is derived to calculate the thermal
expectation value  of the observable $\Omega$. It can be done in EFT-2 formulation  analytically. However,
for the present formulation this process has to be done in a numerical way,
since the matrix representation of the   operators mentioned  above have
dimensions $16\times 16$ and it is impossible to perform the diagonalization
process analytically.

To evaluate the expression Eq. \re{denk3} for the magnetization of the system,
let us start with obtaining the matrix representation of the 4-spin Hamiltonian. Magnetization
per spin in the selected cluster can be calculated by setting $\Omega=S_k^z$ in Eq. \re{denk3},

\eq{denk4}{
m_k=\sandd{S_k^z}=\sandd{\frac{Tr_{4}S_k^z\exp{\paran{-\beta \mathcal{H}^{(4)}}}}{Tr_{4}\exp{\paran{-\beta \mathcal{H}^{(4)}}}}}, \quad k=1,2,3,4.
}

\noindent Let us denote the bases set by $\{\psi_i\}$, where $i=1,2,\ldots 16$. Each of the element of this
bases set can be represented by $\ket{s_1s_2s_3s_4}$, where $s_k=\pm 1 $ is just one spin eigenvalues of the
operator $S_k^z$  ($k=1,2,3,4)$. In this representation, operators in the 4-spin cluster acts on a base via

\eq{denk5}{
\begin{array}{lcl}
S_i^z\ket{\ldots s_i \ldots }&=&s_i\ket{\ldots s_i \ldots }\\
S_i^xS_j^x\ket{\ldots s_i \ldots s_j \ldots }&=&\ket{\ldots -s_i \ldots -s_j \ldots }\\
S_i^yS_j^y\ket{\ldots s_i \ldots s_j \ldots }&=&-s_is_j\ket{\ldots -s_i \ldots -s_j \ldots }\\
S_i^zS_j^z\ket{\ldots s_i \ldots s_j \ldots }&=&s_is_j\ket{\ldots s_i \ldots s_j \ldots },
\end{array}
} where $i,j=1,2,3,4$. Let us denote the matrix representation of the 4-spin Hamiltonian
given in Eq. \re{denk2}  as $H^{(4)}$, which has elements  $H^{(4)}_{ij}=\sand{\psi_i}{\mathcal{H}^{(4)}}{\psi_j}$.
Calculated matrix elements can be found in Appendix \ref{sec_app_a}. The trace of
the exponential of this matrix is just the partition function of the 4-spin cluster.
In order to exponentiate the matrix $H^{(4)}$, it should be diagonalized. The matrix $H^{(4)}$  can be diagonalized
by performing the usual $E^{-1}H^{(4)}E$ similarity transformation, where $E$ is the matrix which has columns as
eigenvectors of the matrix $H^{(4)}$, and $E^{-1}$ is an inverse of it. In the
diagonal form of $H^{(4)}$, the diagonal elements will be the eigenvalues of it.
Let the matrix $H^{(4)}$ be in the diagonal form with the bases set $\{\widetilde{\psi}_i\}$,
then the diagonal elements can be given by
\eq{denk6}{
r_i=\sand{\widetilde{\psi}_i}{\mathcal{H}^{(4)}}{\widetilde{\psi}_i}, i=1,2,\ldots 16
.} Also, let the diagonal elements of the operator  $S_k^z$ in the same bases set be
\eq{denk7}{
t_i^{(k)}=\sand{\widetilde{\psi}_i}{S_k^z}{\widetilde{\psi}_i}, i=1,2,\ldots 16, \quad k=1,2,3,4.
}
The magnetization of the site $k$ can be calculated with writing Eqs. \re{denk6} and \re{denk7}
in Eq. \re{denk4}
\eq{denk8}{
m_k=\sandd{S_k^z}=\sandd{\frac{\summ{i=1}{16}{}t_i^{(k)}\exp{\paran{-\beta r_i}}}{\summ{i=1}{16}{}\exp{\paran{-\beta r_i}}}}, k=1,2,3,4
.}

Aside from the calculation of the term in the thermal average of the right hand side of Eq. \re{denk8},
we are faced with the problem of
taking the thermal average of it. Let us write Eq. \re{denk8} in the closed form as
\eq{denk9}{
m_k=\sandd{f_k\paran{\beta,\{J_i\},\{h_j\}}},
} where $i=1,2,3$ and $j,k=1,2,3,4$. The terms $h_j$ are given by
\eq{denk10}{
\begin{array}{lcl}
h_1&=& J_1^z \paran{S_{11}+S_{12}+S_5}+J_3^zS_6\\
h_2&=& J_1^z \paran{S_{21}+S_{22}+S_9}+J_3^z\paran{S_7+S_8}\\
h_3&=& J_2^z \paran{S_{31}+S_{32}+S_7}\\
h_4&=& J_2^z \paran{S_{41}+S_{42}+S_6}.\\
\end{array}
} Here the spins labeled by $5,6,7,8,9$ are the nearest neighbor of the spins of our 4-spin cluster in
the figure plane and they
can be seen in Fig. \ref{sek1}. The other spins which are labeled by $i1$ and $i2$ are the nearest
neighbor spins of the spin $i$ ($i=1,2,3,4$),  which are belong
to the upper and lower plane, respectively.  With using differential operator technique \cite{ref28}  we can write Eq. \re{denk9} as
\eq{denk11}{
m_k=\sandd{\prodd{j=1}{4}{}\exp{\paran{h_j\nabla_j}}}f_k\paran{\beta,\{J_i\},\{x_j\}}|_{\{x_j\}=0}.
}
Here the functions $f_k\paran{\beta,\{J_i\},\{x_j\}}, (k=1,2,3,4) $ are nothing but the functions given in Eq. \re{denk8} with replecements of all $h_j, (j=1,2,3,4)$ with $x_j, (j=1,2,3,4)$, i.e.

\eq{denk12}{
f_k\paran{\beta,\{J_i\},\{x_j\}}=\left. \frac{\summ{i=1}{16}{}t_i^{(k)}\exp{\paran{-\beta r_i}}}{\summ{i=1}{16}{}\exp{\paran{-\beta r_i}}}
\right|_{h_j\rightarrow x_j}
.}

\noindent In Eq. \re{denk11},  $\nabla_j$  stands for the differential operator with respect to $x_j$ ($j=1,2,3,4$). We note that, we denote the function $f_k\paran{\beta,\{J_i\},\{x_j\}}$ as $f_k\paran{\{x_j\}}$ in the remaining part of the text.   The effect of the exponential differential operator to an arbitrary  function $F(\{x_j\})$ is given by
\eq{denk13}{\exp{\paran{\summ{j=1}{4}{}h_j\nabla_j}}F\paran{\{x_j\}}=F\paran{\{x_j+h_j\}}.}
Let us define the operator
\eq{denk14}{
\theta_{ij}^{(k)}=\exp{\paran{J^{z}_i\nabla_j s_k}}=\left[\cosh{\paran{J_i^z\nabla_j}}+s_k \sinh{\paran{J_i^z\nabla_j}}\right].
}
With writing the terms given in Eq. \re{denk10} in Eq. \re{denk11},  we can write Eq. \re{denk11} with operators defined in
Eq. \re{denk14} as follows:

\eq{denk15}{
m_k=\sandd{\theta_{11}^{(11)} \theta_{11}^{(12)} \theta_{11}^{(5)} \theta_{31}^{(6)} \theta_{12}^{(21)} \theta_{12}^{(22)} \theta_{12}^{(9)} \theta_{32}^{(7)} \theta_{32}^{(8)}
\theta_{23}^{(31)} \theta_{23}^{(32)} \theta_{23}^{(7)}  \theta_{24}^{(41)} \theta_{24}^{(42)} \theta_{24}^{(6)} }f_k\paran{\{x_j\}}|_{\{x_j\}=0}
.}
Writing Eq. \re{denk14} in Eq. \re{denk15} and using decoupling approximation \cite{ref28} will give the expressions of magnetizations
as,

\eq{denk16}{
m_k=\left[\paran{\phi_{11}^{(1)}}^2  \phi_{11}^{(2)} \phi_{31}^{(3)} \paran{\phi_{12}^{(2)}}^2  \phi_{12}^{(1)} \phi_{32}^{(3)} \phi_{32}^{(4)}
\paran{\phi_{23}^{(3)}}^2  \phi_{23}^{(4)}  \paran{\phi_{24}^{(4)}}^2  \phi_{24}^{(3)} \right]f_k\paran{\{x_j\}}|_{\{x_j\}=0}
} where $k=1,2,3,4$ and these new operators in this case defined as

\eq{denk17}{
\phi_{ij}^{(k)}=\left[\cosh{\paran{J_i^z\nabla_j}}+m_k \sinh{\paran{J_i^z\nabla_j}}\right]
.}

With writing Eq. \re{denk17} in Eq. \re{denk16}, we can get the magnetization expressions in a closed form as

\eq{denk18}{
m_k=\summ{p=0}{3}{}\summ{q=0}{3}{}\summ{r=0}{4}{}\summ{t=0}{3}{}C^{(k)}_{pqrt}m_1^pm_2^qm_3^rm_4^t, \quad k=1,2,3,4
.} Here the coefficients $C^{(k)}_{pqrt}$ can be calculated via Eq. \re{denk13}. Since the expansions of Eq. \re{denk18} a bit
longer and complicated, we do not want to give the explicit forms here.
Solving these number of four nonlinear equations given in Eq. \re{denk18}  gives the magnetizations.

Numerical solution process of Eq. \re{denk18} may be Newton-Raphson iteration or similar iterative methods \cite{ref30}. During this iteration process we have to calculate evaluations of the functions defined in \re{denk12} on the $\{x_j\}$ points many times. Indeed, a significant number of these calculations belongs to the same  $\{x_j\}$ points. Since this calculation step consist of the diagonalization process of the $16\times 16$ matrices, then we have to avoid unnecessary  calculation repetitions. For this purpose, we can use an alternative form of the \re{denk18} which can be given by,

\eq{denk19}{
m_k=\summ{i_1=-1,}{1}{}\summ{i_2=-2}{2}{}D_{i}  f_k\left[\paran{p+q}J_1^{(z)}+rJ_3^{(z)},
\paran{s+t}J_1^{(z)}+\paran{v+w}J_3^{(z)},\paran{x+y}J_2^{(z)},\paran{z+o}J_2^{(z)}\right]
.}
The details of the derivation can be found in Appendix \ref{sec_app_b}.
There are number of eleven distinct summations in Eq. \re{denk19}  and index $i$ stands for all of these indexes of summations, while the index $i_1$ is a short notation of the indexes $q,r,t,v,w,y,o$  and  the $i_2$ is short notation of the indexes $p,s,x,z$. The incerement of these indexes in all of  the summations in Eq. \re{denk19}, should be two. The coefficients in Eq. \re{denk19} given by
\eq{denk20}{
D_i=\frac{1}{2^{15}}A_1^{(p,t)}A_2^{(q,s)}A_3^{(r,v,x,o)}A_4^{(w,y,z)}
} where
$$
A_1^{(p,t)}=2_p\paran{1-m_1}^{(3-p-t)/2}\paran{1+m_1}^{(3+p+t)/2},
$$
\eq{denk21}{
A_2^{(q,s)}=2_s\paran{1-m_2}^{(3-q-s)/2}\paran{1+m_2}^{(3+q+s)/2},
}
$$
A_3^{(r,v,x,o)}=2_x\paran{1-m_3}^{(5-r-v-x-o)/2}\paran{1+m_3}^{(5+r+v+x+o)/2},
$$
$$
A_4^{(w,y,z)}=2_z\paran{1-m_4}^{(4-w-y-z)/2}\paran{1+m_4}^{(4+w+y+z)/2},
$$
and where
\eq{denk22}{2_p=\komb{2}{(2-p)/2}=\frac{2!}{\paran{\frac{2-p}{2}}!\paran{\frac{2+p}{2}}!}}
is denotes the combination.

Since all magnetizations are close to zero in the vicinity of the (second order) critical point, we can obtain another coupled
equation system to determine the transition temperature, by linearizing the equation system given in  Eqs. \re{denk18} or \re{denk19}.
If we denote the coefficient matrix of this linear equation system by $A$, then  the critical temperature
can be determined from equation $det(A)=0$, where $det$ stands for the determinant of a matrix.

\section{Results and Discussion}\label{results}

In order to obtain information about the effect of the anisotropy in the exchange interaction on
the critical temperature and the order parameter of the nanotube, it is beneficial to introduce the following definition
\eq{denk23}{
J_1^z=J_2^z=J_3^z=J
} and scale all other components of the exchange interaction with $J$ as,
\eq{denk24}{
r_i=\frac{J_i^x}{J}=\frac{J_i^y}{J}, \quad i=1,2,3.
} It is clear that the value of $r_i=0.0$  corresponds to  Ising nanotube, and as $r_i$ begins
to increase starting from $0.0$ to $1.0$, the model arrive the isotropic Heisenberg model with passing
the anisotropic Heisenberg model by means of the XXZ model. For determining the critical temperature
that separetes the ordered and disordered phases we use the
procedure  mentioned in Sec. \ref{formulation}. The phase transition temperature of the core-shell nanotube
system sensitively depends on the anisotropies in the exchange interaction of the
system i.e. $r_i,(i=1,2,3)$.

\subsection{Phase Diagrams}\label{results_phase}

The critical temperature of the considered system has been found as $k_BT_c/J=5.033$ for highly
anisotropic case $r_1=r_2=r_3=0$. The obtained result is nothing but the critical temperature of
the Ising nanotube and it can be compared with the value of $k_BT_c/J=5.214$ \cite{ref11_ek},
which is the result of the Ising nanotube within the EFT formulation.
Our finding value is slightly lower than that obtained EFT value this is because there
exists some distinct differences between two formulation schemes.

\begin{figure}[h]\begin{center}
\epsfig{file=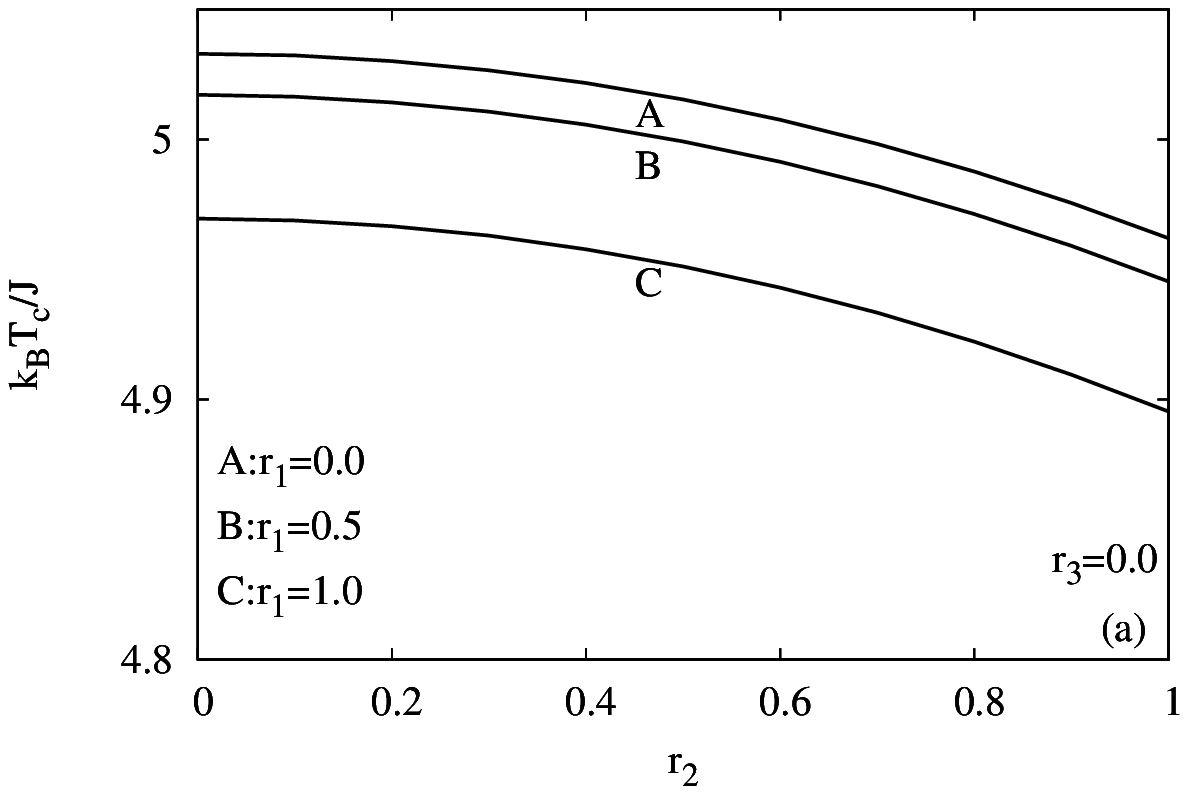, width=6cm}
\epsfig{file=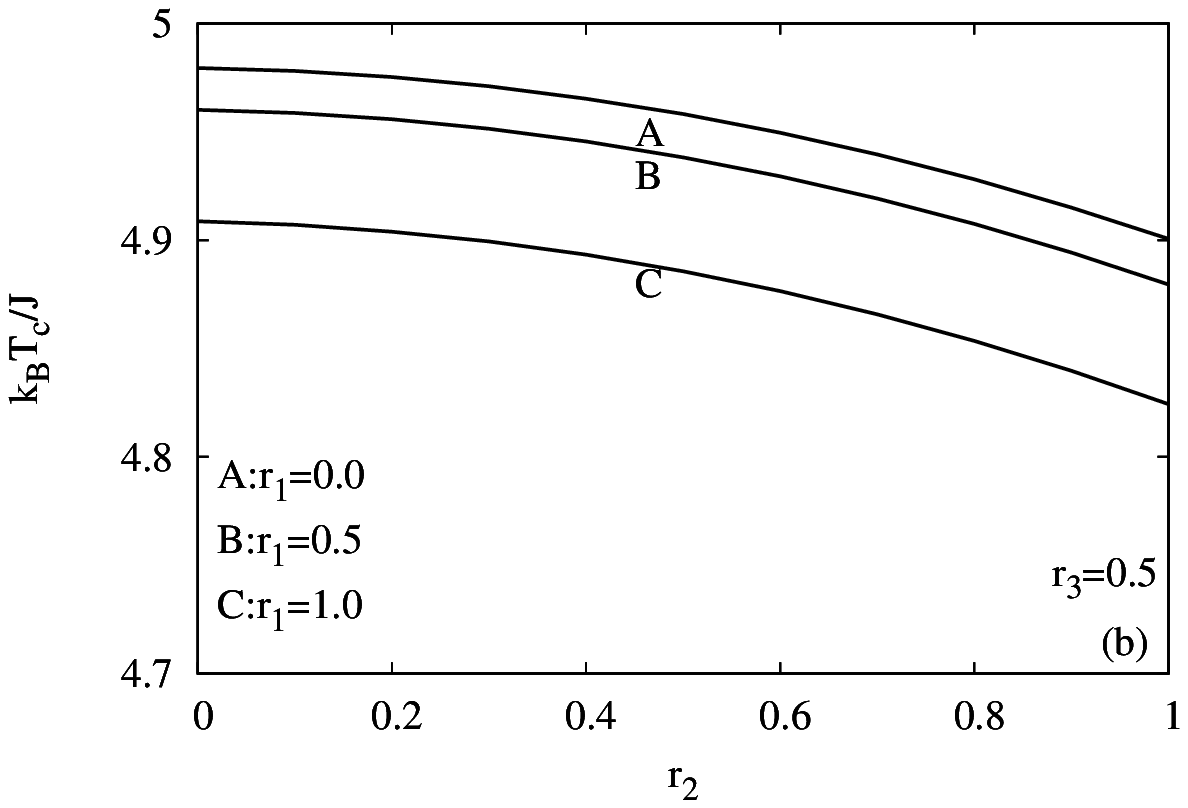, width=6cm}
\epsfig{file=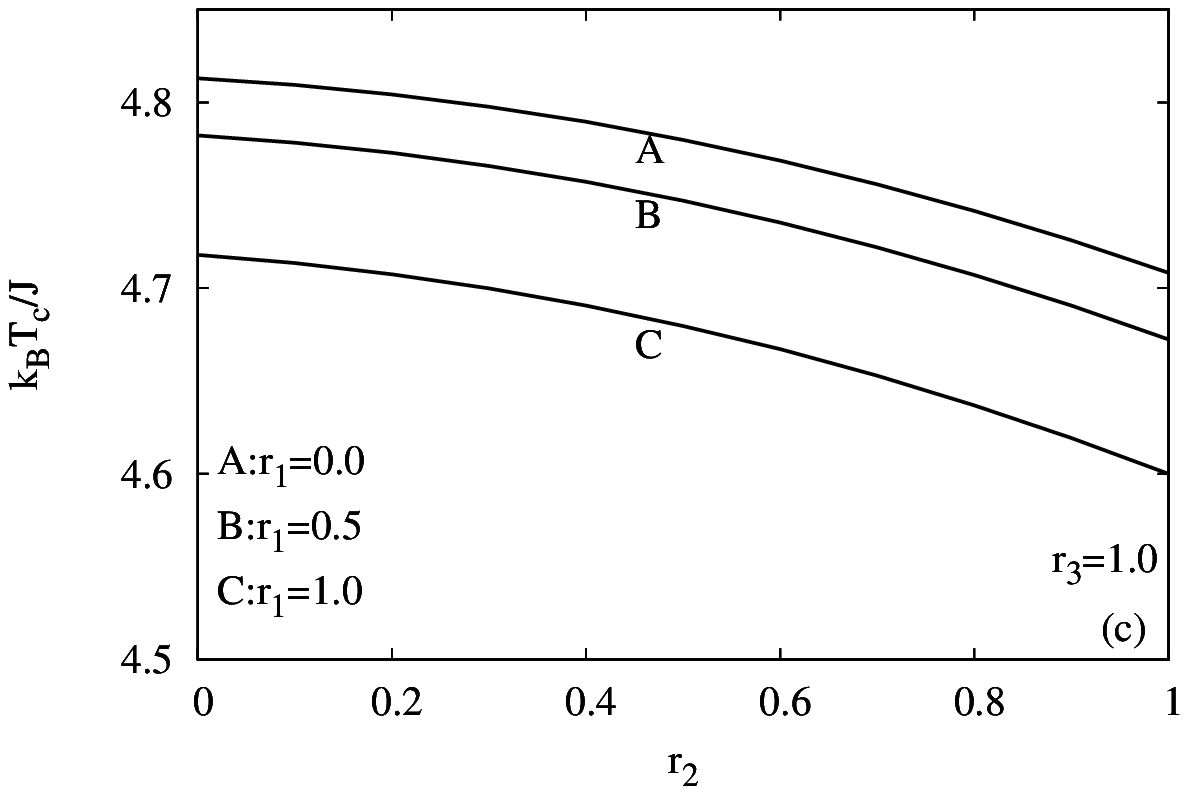, width=6cm}
\end{center}
\caption{The variation of the critical temperature of the Heisenberg nanotube with $r_2$, for selected  values
of $r_1=0.0,0.5,1.0$ and (a) $r_3=0.0$,  (b) $r_3=0.5$,  (c) $r_3=1.0$.} \label{sek2}\end{figure}

In accordance with the expectations when the anisotropy in the exchange interaction
decreases, then the phase transition temperature of the system decreases.
One can clearly see this situation in Fig. \ref{sek2} where we illustrate the phase diagrams of core-shell nanotube system
in temperature and $r_2$ plane for selected values of $r_1,r_3=0.0,0.5,1.0$. It is possible to see some special cases in
constructed diagrams. For example, the core and the interaction of the core and shell
will be Ising type (the curve labeled by A in Fig. \ref{sek2} (a)) in the case of  $r_1,r_3=0.0$ while
for the values of $r_1,r_3=1.0$ these interactions are isotropic Heisenberg type
(the curve labeled by C in Fig. \ref{sek2} (c)), etc. We note that, for the value of  $r_1=r_2=r_3=1.0$ the
critical temperature of the isotropic Heisenberg nanotube is $k_BT_c/J=4.600$.
To the best of our knowledge, this is the first result of the Heisenberg nanotube in the literature.

As seen in Fig. \ref{sek2}, the phase transition temperature of the studied system gradually
decreases when the value of $r_2$ rises starting from zero and the discussed behavior
does not sensitively depend on the values  of $r_1$  and $r_3$. It is obvious that
an increment in value of $r_2$ corresponds to the decreasing anisotropy in the
exchange interaction. At this point, in order to better understanding of
the mechanism underlying the system, it is beneficial to talk about the
limit cases of $r_{2}$ which are 0.0 and 1.0. The first one refers to
the fully anisotropic Ising limit where the spins align in $z$ direction.
The latter case of $r_2$ means the isotropic Heisenberg limit  and
allows spins to align in other direction than in $z$ direction.
The second order phase transition point between ordered and disordered
phases begins to shrink with increasing value of $r_{2}$. In other words, a relatively small
amount of thermal energy is needed to observe a phase transition in the system with
further increment in $r_{2}$.

Moreover, it can be said that all changes of the phase transition temperatures of the core-shell
Heisenberg nanotube system  occur in the range of $(4.600, 5.033)$ for all  combination of
Hamiltoanian  parameters. The lower value of critical  temperature corresponds to the phase
transition temperature of the isotropic Heisenberg nanotube, while the relatively bigger one
refers to the critical temperature of the Ising nanotube. As the value of the $r_i,(i=1,2,3)$ rises,
then critical  temperature of the system decreases within the range mentioned above. In Fig. \ref{sek2}
the labels $A,B$ and $C$ represent the varying $r_{1}$ values. By the way, we want to give point to
the positions of the curves represented in the temperature
and $r_2$ plane. When the interaction type within the core changes from the
Ising type to the $XXZ$ type,  critical temperature of the system decreases for
any value of $r_2$ and $r_3$ (see curves labeled by $B$ and $A$ in Fig. \ref{sek2} (a)-(c)).
Same situation is also valid for the changing of the  interaction type of the core from the $XXZ$ type
to  the isotropic Heisenberg type (see curves labeled by $C$ and $B$ in Fig. \ref{sek2} (a)-(c)).
The physical mechanism underlying this observation is as mentioned above.

\begin{figure}[h]\begin{center}
\epsfig{file=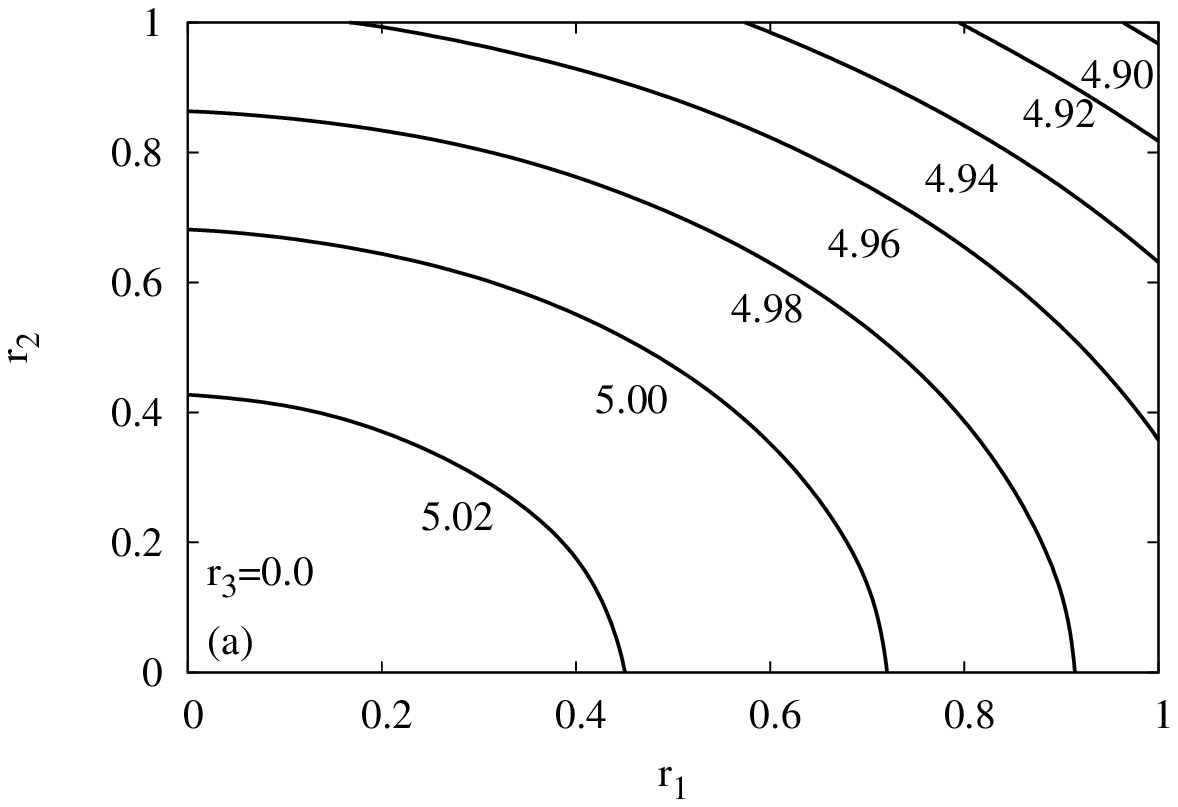, width=6cm}
\epsfig{file=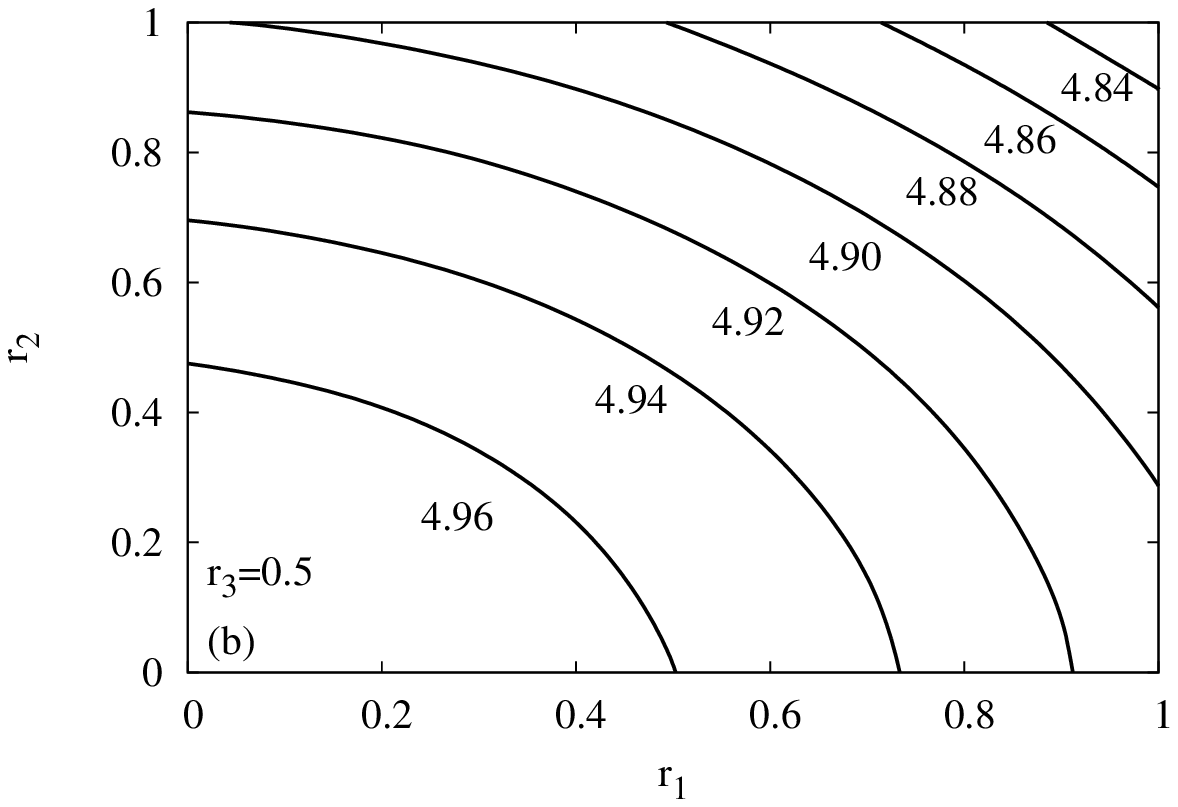, width=6cm}
\epsfig{file=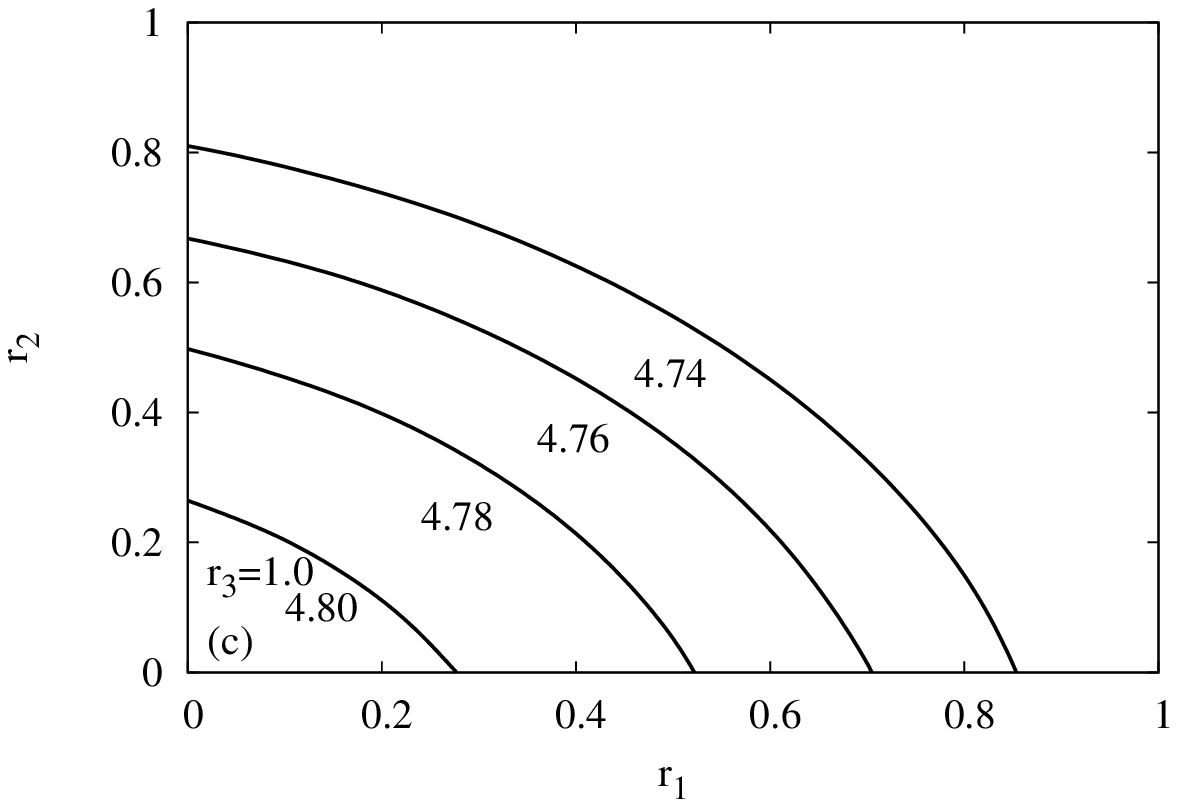, width=6cm}
\end{center}
\caption{The equally valued critical temperature curves of the Heisenberg nanotube in
the $(r_1,r_2)$ plane for (a) $r_3=0.0$,  (b) $r_3=0.5$,  (c) $r_3=1.0$.} \label{sek3}\end{figure}

In the following analysis, in order to see the influences of the anisotropy in the exchange
interaction on the critical temperature of the core-shell nanotube system in another perspective,
the equally valued critical temperature curves in the ($r_1,r_2$) plane are depicted for the
selected values of $r_3=0.0,0.5,1.0$, in Fig. \ref{sek3}. Each numbers accompanying of the curves are the
phase transition temperature and it can clearly deduce from the figure that lower critical temperatures
occur of the considered system for the higher values of $r_1, r_2$. In addition to these, equally valued
critical temperature curves do not exhibit symmetry with respect
to the $r_1=r_2$ line, due to the different number of spins of the core and the shell.
The shell has number of $12$ spins, while the core has number of $6$ spins. It is also expected result
that the critical temperature of the system be more affected by the change of $r_2$ than $r_1$.

\subsection{Behavior of the order parameter}\label{results_mag}

Now, we want to investigate the effect of the anisotropy of the exchange interaction on the order parameter, i.e. the magnetization. To do this, we define the core and the shell magnetization as

\eq{denk25}{
m_c=\frac{m_1+m_2}{2}, \quad m_s=\frac{m_3+m_4}{2}
}
where $m_c$ and $m_s$ stand for the magnetization of the core and  shell, respectively. The values of the $m_i (i=1,2,3,4)$ can be determined with the procedure explained in Sec. \ref{formulation}.

\begin{figure}[h]\begin{center}
\epsfig{file=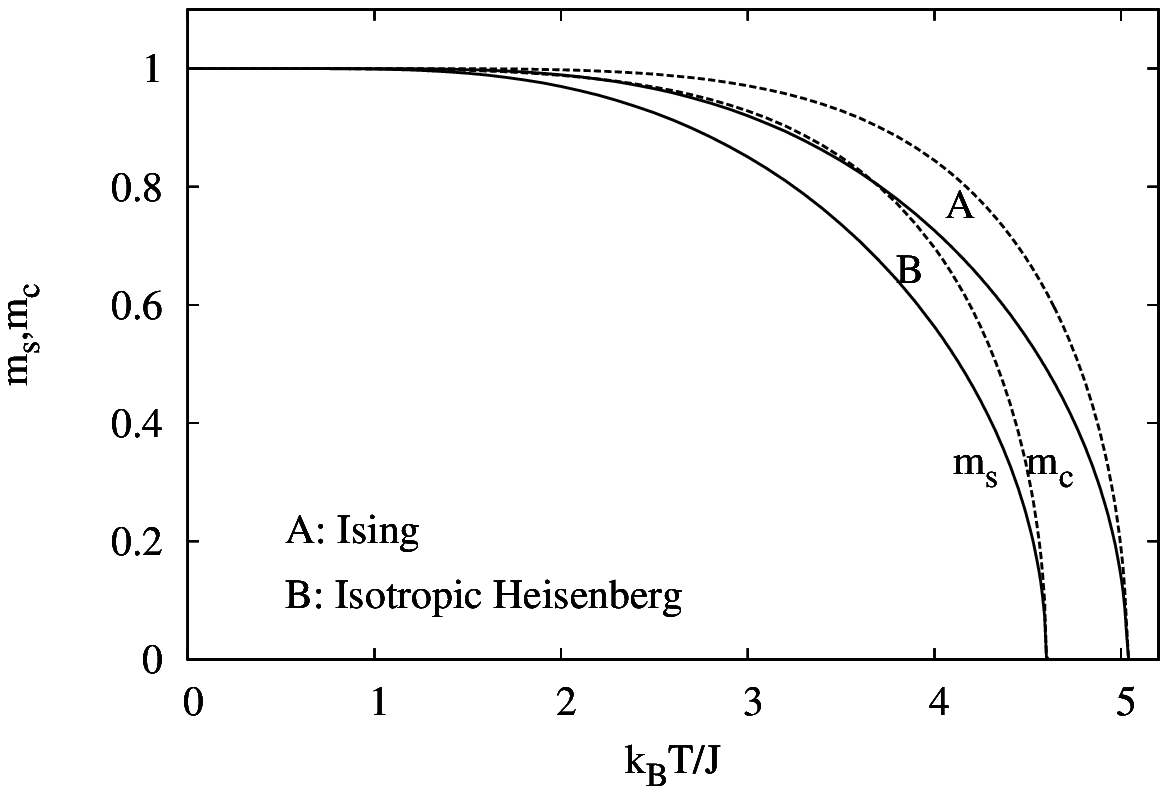, width=12cm}
\end{center}
\caption{Variation of the magnetization of the core ($m_c$) and the shell ($m_s$) with the temperature, for Ising nanotube (labeled by A) and isotropic Heisenberg nanotube (labeled by B). Solid lines represent to the magnetization of the shell, while the dotted lines represent to the magnetization of the core.} \label{sek4}\end{figure}

First, in order to see the difference between the Ising nanotube (i.e. $r_1=r_2=r_3=0.0$) and isotropic Heisenberg nanotube (i.e. $r_1=r_2=r_3=1.0$) we depict the variation of the magnetization of the core and shell with the temperature. This behavior can be seen in Fig. \ref{sek4}.

We can see from the Fig. \ref{sek4} that, for both of the models, the magnetization of the shell lies below of the  magnetization of the core. Note that, the curves labeled by A (i.e. variation of the magnetizaton of the core and shell with temperature for the Ising model) is consistent with corresponding curves in  Ref. \cite{ref10}. The only difference is in the critical temperature (where both of the magnetizations go to zero) and the reason of this point was explained in Sec. \ref{results_phase}. On the other hand the only difference between two model seems about the critical temperatures, when we compare curves labeled by A and B in Fig. \ref{sek4}. We can say that, when we depict the curves correspond to the different anisotropy values, they will be lie between the curves labeled by A and B in Fig. \ref{sek4}.  For this reason let us inspect to excess magnetization of the core which is defined by
\eq{denk26}{
m_x=m_c-m_s
.}

\begin{figure}[h]\begin{center}
\epsfig{file=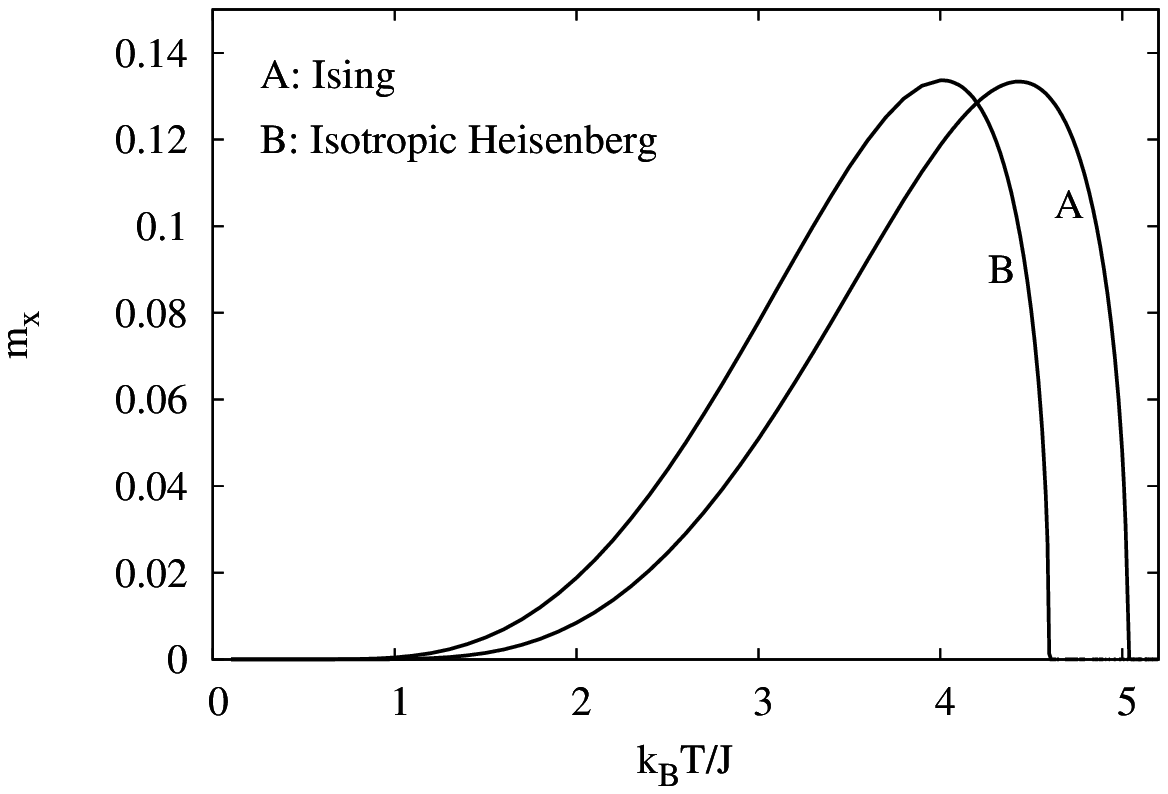, width=12cm}
\end{center}
\caption{Variation of the  excess magnetization of the core ($m_x$, defined in Eq. \re{denk26})  with the temperature for Ising nanotube (labeled by A) and isotropic Heisenberg nanotube (labeled by B).} \label{sek5}\end{figure}

Variation of the excess magnetization of the core with the temperature can be seen in Fig. \ref{sek5} for the Ising nanotube (curve labeled by A) and isotropic Heisenberg nanotube (curve labeled by B).

It can be seen from the  Fig. \ref{sek5} that, for both of the  models, the value of $m_x$ is always greater than the value of zero, or equal to zero, i.e. for both of the models  $m_c\ge m_s$  always satisfied. The behavior of the $m_x$ with the rising temperature is same for both of the model:   stays almost constant value of zero for a while, then rises and decreases sharply. After the related critical temperature $m_x=0.0$, since both of the magnetizations of the core and the shell is zero.  But when we compare both of the curves quantitatively we can see one point. Quantitative relation of  $m_x$ between these two model is: for low temperatures, the nonzero value of $m_x$ for the Ising model is lower than the  value of $m_x$ for  isotropic Heisenberg model. This relation getting reverse after a specific value of the temperature, which is close to the critical temperature of the isotropic Heisenberg model,  while temperature rises.

\section{Conclusion}\label{conclusion}

In this work, EFT-4 formulation for the anisotropic Heisenberg model on the
core-shell nanotube geometry is derived and the variation of the critical  temperature with
the anisotropy in the exchange interaction is obtained. The system is
handled with three exchange interactions, namely $r_i,(i=1,2,3)$,
which controls the anisotropy in the exchange interactions between the
nearest neighbor spins in the core, shell and nearest neighbor spins
belongs to core and shell, respectively.

The formulation is derived for the finite cluster, which has number of four spins.
Due to the impossibility of the obtaining the eigenvalues of the matrix representation
of the 4-spin Hamiltonian, numerical procedures applied. For the diagonalization
of the 4-spin Hamiltonian matrix, Jacobi transformation has been used and then for obtaining
the critical temperature of the system for a given Hamiltonian parameters,
standard LU decomposition has been used.  For interested readers, matrix representation of the
4-spin Hamiltonian has been given in Appendix \ref{sec_app_a}.

In the Ising case (i.e. all exchange interactions in the system are Ising type), the critical temperature of the
nanotube obtained as $k_BT_c/J=5.033$. This value is slightly lower than that of in the literature.
This is due to the formulation used here. EFT-4 formulation solves the system with using larger cluster
and it is believed that, using larger clusters gives more accurate results. In the other extreme case, namely
isotropic Heisenberg nanotube, the critical temperature obtained as $k_BT_c/J=4.600$. As far as we know,
this is the first result for the Heisenberg model on the nanotube geometry in the literature.
Changing anisotropy in the exchange interaction, changes the critical temperature of the
nanotube within this two special cases.

On the other hand, the effect of the anisotropy in the exchange interaction on the order parameter (magnetization) has been investigated.
Behavior of the magnetization with rising temperature is typical both limits of the system, i.e. for Ising model and isotropic Heisenberg model. The ground state magnetization does not affected by the rising anisotropy in the exchange interaction. The only change occurs in the value of tempeature that the magnetization goes to zero, i.e. critical temperature. Besides, the difference between the magnetization of the core and the shell has been investigated. It has been found that, at a lower temperatures the magnetization of the core and shell are equal to each other, regardless of the model (Ising, XXZ type anisotropic Heisenberg or isotropic Heisenberg). When temperature rises, the magnetization of the core takes the higher values than the magnetization of the shell, due to the excess interaction per spin at the core. The last observation on the behavior of the order parameter was, the difference of the magnezition values between the core and shell getting bigger in the intermediate (far from zero and the critical temperatures of the related model) temperatures when the anisotropy in the exchange interaction lowers.

We hope that the results  obtained in this work may be beneficial form both theoretical and
experimental point of view.

\appendix
\section{Matrix Representation of the 4-spin Hamiltonian}\label{sec_app_a}

The matrix representation of Eq. \re{denk2} symmetric and  it has number of $96$ nonzero elements. This matrix representation can be given with number of $22$ distinct elements as follows:
\begin{changemargin}{-5cm}{-3cm}
\eq{denk_app_1}{\paran{
\begin{array}{cccccccccccccccc}
K_{1}&0&0&0&0&K_{22}&K_{20}&K_{20}&K_{20}&0&K_{21}&0&0&0&0&0\\
&K_{2}&K_{18}&K_{17}&K_{17}&0&0&0&0&0&0&K_{20}&0&K_{21}&0&0\\
&&K_{3}&K_{17}&0&0&0&0&0&0&0&K_{20}&K_{20}&0&K_{21}&0\\
&&&K_{4}&K_{19}&0&0&0&0&0&0&K_{9}&0&K_{20}&0&0\\
&&&&K_{5}&0&0&0&0&0&0&0&K_{9}&K_{20}&K_{20}&0\\
&&&&&K_{6}&K_{17}&0&K_{17}&K_{17}&0&0&0&0&0&K_{21}\\
&&&&&&K_{7}&K_{19}&K_{18}&0&K_{17}&0&0&0&0&0\\
&&&&&&&K_{8}&0&K_{18}&K_{17}&0&0&0&0&K_{20}\\
&&&&&&&&K_{9}&K_{19}&0&0&0&0&0&K_{20}\\
&&&&&&&&&K_{10}&K_{17}&0&0&0&0&K_{20}\\
&&&&&&&&&&K_{11}&0&0&0&0&K_{22}\\
&&&&&&&&&&&K_{12}&K_{19}&0&K_{17}&0\\
&&&&&&&&&&&&K_{13}&K_{17}&K_{17}&0\\
&&&&&&&&&&&&&K_{14}&K_{18}&0\\
&&&&&&&&&&&&&&K_{15}&0\\
&&&&&&&&&&&&&&&K_{16}\\
\end{array}
}
}
\end{changemargin}
where the terms $K_i$ ($i=1,2,\ldots, 22$) defined by

\eq{denk_app_2}{
\begin{array}{lcl}
K_{1}&=&-J_1^z-J_2^z-3J_3^z-h_{1}-h_{2}-h_{3}-h_{4}\\
K_{2}&=&J_1^z-J_2^z+J_3^z+h_{1}-h_{2}-h_{3}-h_{4}\\
K_{3}&=&J_1^z-J_2^z-J_3^z-h_{1}+h_{2}-h_{3}-h_{4}\\
K_{4}&=&-J_1^z+J_2^z+J_3^z-h_{1}-h_{2}+h_{3}-h_{4}\\
K_{5}&=&-J_1^z+J_2^z-J_3^z-h_{1}-h_{2}-h_{3}+h_{4}\\
K_{6}&=&-J_1^z-J_2^z+3J_3^z+h_{1}+h_{2}-h_{3}-h_{4}\\
K_{7}&=&J_1^z+J_2^z+J_3^z+h_{1}-h_{2}+h_{3}-h_{4}\\
K_{8}&=&J_1^z+J_2^z-J_3^z+h_{1}-h_{2}-h_{3}+h_{4}\\
K_{9}&=&J_1^z+J_2^z-J_3^z-h_{1}+h_{2}+h_{3}-h_{4}\\
K_{10}&=&J_1^z+J_2^z+J_3^z-h_{1}+h_{2}-h_{3}+h_{4}\\
K_{11}&=&-J_1^z-J_2^z+3J_3^z-h_{1}-h_{2}+h_{3}+h_{4}\\
K_{12}&=&-J_1^z+J_2^z-J_3^z+h_{1}+h_{2}+h_{3}-h_{4}\\
K_{13}&=&-J_1^z+J_2^z+J_3^z+h_{1}+h_{2}-h_{3}+h_{4}\\
K_{14}&=&J_1^z-J_2^z-J_3^z+h_{1}-h_{2}+h_{3}+h_{4}\\
K_{15}&=&J_1^z-J_2^z+J_3^z-h_{1}+h_{2}+h_{3}+h_{4}\\
K_{16}&=&-J_1^z-J_2^z-3J_3^z+h_{1}+h_{2}+h_{3}+h_{4}\\
K_{17}&=&-J_3^y-J_3^x\\
K_{18}&=&-J_1^y-J_1^x\\
K_{19}&=&-J_2^y-J	_2^x\\
K_{20}&=&J_3^y-J_3^x\\
K_{21}&=&J_2^y-J_2^x\\
K_{22}&=&J_1^y-J_1^x.\\
\end{array}
}

Here, the following bases set ($\ket{\psi_i}=\ket{s_1s_2s_3s_4}, i=1,2,\ldots, 16$) has been used:

\eq{denk_app_3}{
\begin{array} {ccccccc}
\ket{\psi_1}&=&\ket{1111}&,&\ket{\psi_9}&=&\ket{1-1-11},\\
\ket{\psi_2}&=&\ket{-1111}&,&\ket{\psi_{10}}&=&\ket{1-11-1},\\
\ket{\psi_3}&=&\ket{1-111}&,&\ket{\psi_{11}}&=&\ket{11-1-1},\\
\ket{\psi_4}&=&\ket{11-11}&,&\ket{\psi_{12}}&=&\ket{-1-1-11},\\
\ket{\psi_5}&=&\ket{111-1}&,&\ket{\psi_{13}}&=&\ket{-1-11-1},\\
\ket{\psi_6}&=&\ket{-1-111}&,&\ket{\psi_{14}}&=&\ket{-11-1-1},\\
\ket{\psi_7}&=&\ket{-11-11}&,&\ket{\psi_{15}}&=&\ket{1-1-1-1},\\
\ket{\psi_8}&=&\ket{-111-1}&,&\ket{\psi_{16}}&=&\ket{-1-1-1-1}.
\end{array}
}

In order to obtain Eq. \re{denk8}, the matrix given in Eq. \re{denk_app_1} has been numerically diagonalized throughout the calculations.

\section{Derivation of Eq. \re{denk19}}\label{sec_app_b}

In order to derive the Eq. \re{denk19} let us start with one spin cluster. Let a spin in the system has number of $z_1$ nearest neighbor and let only exchange interaction in the system be $J_1^z$. Suppose that, the operator given in Eq. \re{denk16} be for this one spin cluster as

\eq{denk_app_c_1}{
m_1=\left[\paran{\phi_{11}^{(1)}}\right]^{z_1}f_1\paran{x_1}|_{x_1=0}
}
where the definition of the operator is given in Eq. \re{denk17}.

Let us write \re{denk_app_c_1} with converting hypertrigonometric functions in \re{denk17} to exponentials
\eq{denk_app_c_2}{
m_1=\frac{1}{2^{z_1}}\left[\paran{1+m_1}\exp{\paran{J_1^z\nabla_1}}+ \paran{1-m_1}\exp{\paran{-J_1^z\nabla_1}}\right]^{z_1}f_1\paran{x_1}|_{x_1=0}
.} With expanding the right hand side of Eq. \re{denk_app_c_2} with Binomial expansion, then  applying Eq. \re{denk13},
we can arrive the equality
\eq{denk_app_c_3}{
m_1=\frac{1}{2^{z_1}}\summ{n=0}{z_1}{}\komb{z}{n}\paran{1+m_1}^{z_1-n}\paran{1-m_1}^{n} f_1\left[\paran{z_1-2n}J_1^z\right]
.}
By the change of variable $t=z_1-2n$ in Eq.  \re{denk_app_c_3} we get the expression
\eq{denk_app_c_4}{
m_1=\frac{1}{2^{z_1}}\summ{t=-z_1}{z_1}{}\komb{z_1}{(z_1-t)/2}\paran{1+m_1}^{(z_1+t)/2}\paran{1-m_1}^{(z_1-t)/2} f_1\left(t J_1^z\right)
}
and this completes the derivation for one spin cluster.

Eq. \re{denk_app_c_4} can be easily generalized to the clusters, that have more than one spin.

\end{document}